# Structural, Electromagnetic, and Thermoelectric properties of $Bi_4O_4S_3$ Superconductor

P. Srivatsava, Shruti, and S. Patnaik
School of Physical Sciences, Jawaharlal Nehru University, New Delhi-110067, India.

## Abstract

We report on the synthesis and extensive characterization of layered $Bi_4O_4S_3$ superconductor. This is the optimally doped sample with $T_c$ ~5.3 K out of a series of $Bi_6O_4S_4(SO_4)_{1-x}$ samples synthesized by solid state reaction. The series was synthesized towards establishing a phase diagram of transition temperature as a function of carrier concentration. Crystal structure for $Bi_4O_4S_3$ shows larger bending of Bi2-S2-Bi2 bond in the $BiS_2$ layer compared to that for the parent phase. Scanning electron microscopy images show platelets like morphology for $Bi_4O_4S_3$ signifying the layered structure. While the parent compound is found to be semiconducting, the electrical resistivity of $Bi_4O_4S_3$ exhibits $T^2$ dependence in a small temperature range between 25 and 50 K. The typical dome structure for variation of $T_c$ with dopant concentration is not observed. From the magneto-transport data $H_{c2}$ for $Bi_4O_4S_3$ is estimated to be ~2.75T with WHH approximation and the corresponding coherence length is ~110Å. Support for multiband signatures is not seen from magneto-resistance data. RF susceptibility data fits well for S-wave isotropic gap with gap value higher than BCS strong coupling limit. Hall measurements confirm dominance of electronic transport with charge carrier density of $4.405\times10^{19}$ $cm^{-3}$ at 10 K. The experimental value of Seebeck coefficient at 35K is well in accord with the calculated value deduced by using density of charge carriers from Hall experiments. The Sommerfeld constant $\gamma$ is estimated to be ~ 1.113 mJ/$K^2$ mol. Evidence for thermally activated flux flow is observed and the pinning potential is found to scale as $B^{-0.3}$ for B < 0.1T and $B^{-1.99}$ for B > 0.1T.

e-mail: spatnaik@mail.jnu.ac.in

## Introduction

The recent discovery of superconductivity in bismuth oxy-sulphide $Bi_4O_4S_3$ has underlined the serendipitous possibilities in layered chalcogenides.[1-6] Pure bismuth itself has five different phases that are superconducting below 8.5 K, therefore the initial challenges focussed on ruling out impurity phases and establish bulk superconductivity with $T_c \sim 4.4$ K.[2] It was quickly deciphered that induction of superconductivity in this compound of tetragonal crystal structure is achieved through $SO_4$ deficiency in the parent $Bi_6O_8S_5$.[1,2] It is now established that like $CuO_2$ and FeAs layers of high $T_c$ cuprates and pnictides, superconductivity in this oxy-sulphide is confined to $BiS_2$ layers. Subsequently, following ideas from oxypnictides, $LaOBiS_2$ is traced to be superconducting when doped with F at the place of O.[7-15] Superconductivity is also confirmed when La is replaced by the rare earth elements Nd, Ce, Pr and Yb with corresponding $T_c$ = 5.6[16](5.2K),[17] 3.0,[18] 5.5 K[19] and 5.4 K[9] respectively. It is suggested that in $YbO_{0.5}F_{0.5}BiS_2$ superconductivity coexists with magnetism below ~2.7 K.[9] It may be noted that Sm/Gd based $BiS_2$ (that shows high $T_c$ in oxypnictides) has not yet been reported even though smaller rare-earth ion Yb based compound is superconducting. It was also suggested that $SrFBiS_2$ might also be a potential family for $BiS_2$ superconductors.[20] This is experimentally verified with recent report of superconductivity in $Sr_{1-x}La_xFBiS_2$.[21] Further, Ag and Cu have also been doped at the place of Bi but in both the cases suppressed superconductivity is reported.[22,23]

Theoretical studies have surmised that the dominating bands for the electron conduction in $Bi_4O_4S_3$ are derived from the Bi $6p_x$ and $6p_y$ orbitals.[1] Strong nesting of the Fermi surface and quasi-one-dimensional bands are also predicted.[24,25] Structurally, $Bi_4O_4S_3$ is composed of a set of $BiS_2$ layers sandwiched between adjacent $Bi_4O_4(SO_4)$ spacer layers.[1] A schematic view of the crystal structure is shown in Figure 1. The Bi atoms are coordinated by four S atoms in the $BiS_2$ plane which form a square-net perpendicular to c-axis. Similar structure is also ascertained for parent phase $Bi_6O_8S_5$. It is inferred, on the basis of band calculations, that $Bi_6O_8S_5$ is an insulator.[1,24] Superconductivity in $Bi_4O_4S_3$ is induced by electron doping into the $BiS_2$ layers via the defects of the $SO_4$ ions at the interlayer site[1] and in particular $Bi_4O_4S_3$ possess 50% defects at a $SO_4$ site. A recent study on Hall and magneto-resistance measurements predicts an exotic multi-band features for $Bi_4O_4S_3$.[3] Further, on the basis of electron-phonon coupling constant calculated for $LaO_{1-x}F_xBiS_2$ ($\lambda = 0.85$), there are indications of a strongly coupled electron-phonon superconductor.[9,10] There are reports on possibility of spin fluctuations leading to unconventional pairing mechanism as evidenced in cuprates, and pnictides.[8] Some theories predict structural instabilities[15,26] which places this system in close proximity to competing ferroelectric and charge density wave (CDW) phases.[15] This instability is possibly caused by anharmonic potential of the S ions lying on the same plane as the Bi ions. Just like the spin-density wave (SDW) instability in Fe-based superconductors, a CDW instability from negative phonon modes at or around the M point, ($\pi$, $\pi$), is suggested to be essential to the superconductivity.[15] As far as pairing symmetry is concerned, three possibilities, *d*-wave, extended *s*-wave, and *p*-wave have been studied based on two orbital RPA calculations.[27]

In a sense, the whole story of $BiS_2$ based layered superconductors is evolving in the side lines of the path taken by oxypnictides. But there are several points where strong departure from pnictides can be observed in disulphides. The broad questions about the multi-band behaviour, BCS coupling parameters and pairing symmetry in ($Bi_4O_4S_3$) the parent superconducting compound have remained unresolved. In this article we report, the evolution of superconducting properties in $Bi_4O_4S_3$ with doping and address these issues along with extensive electromagnetic and thermodynamic characterizations. Towards synthesizing $Bi_4O_4S_3$, we have gradually created deficiency of $SO_4$ in the $SO_4$ layer of the parent compound $Bi_6O_8S_5$ ($Bi_6O_4S_4(SO_4)$. Although a perfect dome has not been observed, we have successfully synthesized $Bi_6O_4S_4(SO_4)_{1-x}$ superconductors. We could achieve the highest $T_c$ phase achieved thus far (for x = 0.5) with $T_c = 5.3$ K. The main focus of the present study is the detailed characterization of $Bi_6O_4S_4(SO_4)_{1-x}$ with nominal composition of x =0 and 0.5. The thermodynamic and transport properties of this material are then studied in detail using an array of experimental techniques such as the Hall Effect, rf susceptibility, electrical resistivity, magento-transport and Thermoelectric power.

**Experimental**

A series of polycrystalline samples of $Bi_6O_4S_4(SO_4)_{1-x}$ (x=0, 0.2, 0.25, 0.4, 0.5, 0.6, 0.66, 0.75, 0.8) were prepared using the conventional solid state reaction method. For synthesis, $Bi_2S_3$, $Bi_2O_3$ powders and S grains were ground, pelletized, sealed into an evacuated quartz tube and heated at 510ºC for 12 h. The product was well-ground, pelletized and again annealed at 510ºC for 12 h in an evacuated quartz tube. Then it was cooled down slowly to room temperature. The second step was repeated for achieving good homogeneity. The resultant sample looked black and hard. The samples were cut to rectangular shape (length 3 mm, width 2.5 mm and thickness of 0.5 mm). Since it was hard, we polished the sample with very fine sand paper and obtained shiny and mirror like surface. The crystallinity of the sample was checked by using the x-ray diffraction (XRD) in a PANanalytical diffractometer with Cu $K_\alpha$ radiation. The analysis of XRD data were done with the Fullprof software. Resistivity and Hall measurements were carried out in low temperature and high magnetic field (8T) Cryogen Free system in conjunction with variable temperature insert. The thermoelectric power (TEP) was measured from 300K to 12K using an automated setup that was attached to a Closed Cycle Refrigerator. A rectangular sample was mounted between two copper blocks and two small chip heaters (50Ω) were used to produce heat gradient. The rf susceptibility was performed by tunnel diode oscillator technique. The sample was kept inside an inductor that formed a part of an LC circuit of an ultrastable oscillator (f~2.3 MHz).

**Results and Discussion**

Figure 2 shows the Rietveld refined XRD patterns for $Bi_4O_4S_3$. The calculated values for lattice parameters, bond lengths and bond angles for the parent $Bi_6O_8S_5$ and superconducting $Bi_4O_4S_3$ samples are given in Table 1. Rietveld refinement of the XRD pattern was carried out using the reported[1,2] Wyckoff positions and space group I4/mmm. Some impurities of Bi (JCPDS No. 05-0519) and $Bi_2S_3$ (JCPDS No. 17-0320, ref. 28) are present which have also been taken into account for refinement that improved the fitting. The quality of the refinement is indicated by the value of $\chi^2$ = 1.75 and 1.68 for the parent and superconducting phase respectively. The refined lattice constants of $Bi_4O_4S_3$ are a = 3.967(5) Å and c = 41.323(1) Å which are slightly less than the values reported earlier[2] and for $Bi_6O_8S_5$ the values are a = 3.965(5) Å and c = 41.234(0) Å. The c axis increased for the superconducting phase indicating relaxation in the structure after losing 50% of $SO_4$ ion from $SO_4$ layer. This is in consonance with decreased $T_c$ with increasing external pressure.[4] The crystal structure made by refined Wyckoff positions indicates that the S and Bi atoms do not stay in a single plane in the $BiS_2$ layer (Figure 1) for the superconducting $Bi_4O_4S_3$ phase. For $Bi_4O_4S_3$, the out of plane angle between S2-Bi2-S2 is found to be 162° and the in-plane S2-Bi2-S2 angle is 88.6° for the adjacent $BiS_2$ layers (Figure 1). This indicates slight deformation from the ideal square lattice of Bi2-S2 atoms. On the other hand, in the parent $Bi_6O_8S_5$ sample, the out of plane S2-Bi2-S2 angle is ~171° and in plane S2-Bi2-S2 angle is 89.7°. This suggests larger bending of Bi2-S2-Bi2 bond in case of superconducting sample.

To investigate the surface morphology of the as grown samples, in Figure 3 we show Scanning Electron Microscopy (SEM) of the parent and superconducting samples. The formation of platelet like morphology for $Bi_4O_4S_3$ is well evident in Figure 3b which is an indication of layered structure. The size of these crystals is in micrometer range having the average length and breadth ~ 6 µm and ~1 µm respectively. Comparing Figure 3a with Figure 3b, we find the parent $Bi_6O_8S_5$ has rather small sized crystals compared to $Bi_4O_4S_3$. This suggests better crystal growth characteristics for the superconducting sample.

Figure 4a show the temperature dependence of resistivity from 1.6 K to 300 K for parent $Bi_6O_8S_5$ (inset i) and 50% $SO_4$ deficient superconducting $Bi_4O_4S_3$. The undoped sample shows semiconducting behaviour up to the lowest measured temperature. The value for room temperature resistivity is ~1 Ohm-cm that reaches to ~36 Ohm-cm at the lowest measured temperature. On the other hand, the electrical resistivity of $Bi_4O_4S_3$ shows metallic behaviour in the normal state. The inset (ii) of Figure 4a shows the resistivity behaviour near the superconducting transition. The onset of transition in this sample is $T_c^{onset}$ = 5.3K as determined by the intersection of extrapolated normal state resistivity line with superconducting transition line. Similarly $T_c$ offset is determined from the intersection at the zero resistance line and is estimated to be $T_{c0}$ ~4.8 K (inset ii, Fig 4a). These values are ~20% higher than earlier reports for this material.[1-6] This could be explained by the fact that with variation in doping concentration we could achieve the optimal doping phase. We further note that the value of resistivity changes from 1.59 mOhm-cm at 300K to 0.786 mOhm-cm just above the transition temperature. Hence, the residual resistivity ratio (RRR) in this case is 2.023. The transport in this polycrystalline sample would have contribution from disorder and perhaps small amounts of impurities. The transition widths for $Bi_4O_4S_3$ $\Delta T_c = T_c(\text{onset}) - T_c(\text{offset})$ = 0.44 K is indicative of excellent grain connectivity. To gain further insight into the normal state electronic correlations, we fit the low-temperature resistivity data to a power law, $\rho = \rho_0 + AT^{\alpha}$, and find that above 50 K, the resistivity data can be reasonably fit by with $\alpha$ = 0.78 where the values of $\rho_0$ and A are estimated to be 0.695 mΩ-cm and 0.01 mΩ-cm/$K^{\alpha}$ respectively (Figure 4a). The data between 25 K and 50 K can be scaled with $\alpha = 2$ (Figure 4b), with the residual resistivity $\rho_0$ is 0.847 mΩ-cm, and the coefficient $A = 3.12\times10^{-5}$ mΩ-cm/$K^2$. The value of $\rho_0$ is large due to the polycrystalline nature of the sample. Below 25K, the resistivity starts deviating from this Fermi liquid fitting. It is to be noted that the value of A is close to the value reported for LaOFeAs.[29] But a clear departure from Fermi liquid behavior near $T_c$ is indicated. In order to get the transition temperature with varying concentration of charge carriers the $T_c$ of $SO_4$ deficient $Bi_6O_4S_4(SO_4)_{1-x}$ samples were also measured. The results have been summarized in inset iii of Fig 4a. It can be seen that superconductivity is not observed upto 25% of $SO_4$ deficiency and decrease in $T_c$ is seen beyond x = 0.5. But superconductivity persists upto the maximum doping that we could achieve (x = 0.8). This doping dependence of $T_c^{onset}$ shows that x = 0.5 is the optimal doping concentration for which transition temperature is highest in $Bi_6O_4S_4(SO_4)_{1-x}$ series. However, we note that no clear dome structure is obtained putting question marks on the possibility of induction of superconductivity onto a strongly correlated insulator.

The accompanying diamagnetic transition is confirmed by measurement of rf susceptibility (rf penetration depth). The data are shown in the inset (ii) of Figure 4b. The measured quantity is the shift in the resonant frequency, $\Delta f \equiv f(T) - f_0 = -4\pi\chi(T)G$, where $\chi$ is the total magnetic susceptibility and $G \simeq f_0V_s/2V_c\,(1 - N)$ is a geometric calibration factor defined by the coil ($V_c$), and the sample ($V_s$) volumes and the demagnetization factor N.[30] We note that magnetic transition of 5.1 K is lower by 0.2 K compared to onset resistivity decrease. As reported elsewhere, detailed analysis of the penetration depth data are consistent with a S-wave gap of $\Delta_0/k_BT_c \sim 2.4$.[31] This decidedly puts the nature of the superconductivity into strong coupling regime.

Figure 5a shows the resistive transition for $Bi_4O_4S_3$ as a function of magnetic field. The $T_c(H)$ values for ρ onset and offset as a function of magnetic field and temperature are shown in inset of Figure 5b, which are equivalently the upper critical $H_{c2}(T)$ and irreversibility ($H_{irr}$ (T)) fields respectively. The slope is estimated to be $[dH_{c2}/dT]|_{T=Tc}$ = −0.75 T/K for $\rho_{onset}$ and −0.45 T/K for $\rho_{offset}$ for H≥ 0.05T. In the conventional Ginzburg – Landau picture, $H_{c2}$ is linear in T near $T_{c0}$ and saturates in the 0K limit. However, in the presence of impurity scattering[32] or in a multiband scenario it shows deviations from the expected behaviour. Nonetheless for the present case, we apply the Werthamer–Helfand–Hohenberg (WHH) equation which relates the slope for $\rho_{onset}$ to $H_{c2}(0) = -0.693T_c[dH_{c2}/dT]|_{T=Tc}$ = 2.754 T. The extrapolated lines in Figure 5b are drawn by using the Ginzburg–Landau equation $H_{c2}(T) = H_{c2}(0)\ (1 - t^2)/(1 + t^2)$ where $t = T/T_c$ is the reduced temperature and $H_{c2}(0)$ is the upper critical field at zero temperature which give the value slightly higher than that by WHH approach. Assuming $H_{c2}(0) = H_{c2}(WHH_0)$, the Ginzburg–Landau coherence length $\xi_{GL}= (\Phi_0/2\pi H_{c2})^{1/2}$, where $\Phi_0$=2.07×10$^{-7}$ G cm$^2$ yields zero temperature coherence length $\xi_{GL}(0)$ ~ 110Å. Further, as seen in, inset of Figure 5a, we note that although bulk superconductivity can be suppressed above 3T, a kink in resistivity is present on the ρ versus T curve. Even with the maximum applied magnetic field of 4T this feature stays at about 5K. In an earlier study, this behaviour has been interpreted as the residual Cooper pairs existing in the system even when the bulk superconductivity has been completely suppressed which supports the applicability of strong coupling theories.[3]

In Figure 6 we show the Hall resistivity $\rho_{xy}$ with varying field at 10K and 300K. The Hall resistivity is negative for all fields and show nonlinearity at both the temperatures which suggests possible deviations from single-band analysis.[2,3,18] Clearly, the conduction mechanism is dominated by electronic transport. The Hall coefficient is field-dependent and the carrier concentration estimated from low field value is n~ 4.405×10$^{19}$ cm$^{-3}$ at 10 K and that increases to ~7.28×10$^{19}$ cm$^{-3}$ at 300 K. These values are almost three times higher than the values reported by Singh et al (~1.53×10$^{19}$ cm$^{-3}$ at 10K) while close to the values given by Li et al (~$4.5\times10^{19}$ $cm^{-3}$ at 2K).[2,3] Moreover, the magnetoresistance for a multiband system is expected to follow $\Delta\rho_{xx}/\rho_0 \propto H^2$ behavior in the low field region and as plotted, in the inset of Figure 6, no such dependence is seen. Clearly more studies are needed to establish the applicability of exotic multi band theories.[3]

The Seebeck coefficient (S) of $Bi_4O_4S_3$ with respect to copper is shown in Figure 7. In agreement with Hall data, S is negative in the whole temperature range varying from ~ −23 μV/K at 300 K to a maximum negative value of ~ −30.5 μV/K at ~35 K. In general, the *S*(T) would consist of three parts; the diffusion term, the spin-dependent scattering term and the phonon-drag term due to electron-phonon interaction.[33] The spin-dependent contribution in particular, would be weak since there are no magnetic elements in $Bi_4O_4S_3$. The phonon-drag effect usually gives a peak structure in the temperature dependence of thermopower (TEP). At high temperature, the phonon-drag part of the TEP varies as 1/T, and at low temperature it goes as ~$T^3$ and consequently there is a peak in S(T) at ~ $\theta_D$ /5 (where $\theta_D$ is the Debye temperature ) in transiting from the high-temperature to the low-temperature regime. Heat capacity measurement done by Takatsu et al. estimates $\theta_D$ ~192 K for $Bi_4O_4S_3$.[5] The pronounced peak in our S(T) data around 35K is very close to the value ~ $\theta_D/5$ = 38 K that confirms the contribution due to the phonon-drag effect. This peak structure was not observed in the in previous reports of TEP in $Bi_4O_4S_3$.[6, 22] Further, the carrier concentration from low-temperature Hall data can be used to obtain a qualitative estimate of diffusive part of Seebeck coefficient and so the effective mass of charge carriers m* and the Sommerfeld coefficient γ can be estimated. If we assume that the electron carrier concentration below ~35 K is close to the value inferred from the Hall data ($4.405\times10^{19}$ electrons/$cm^3$ at 10K), the free electron model predicts a Fermi energy (temperature) of 45.53 meV (528.29 K).[34] In low carrier concentration metals,[33] the Seebeck coefficient is given by the diffusion contribution of the Mott expression or $S = \pi^2 k_B T(2eT_F)^{-1}$, where $T_F$ is the Fermi temperature. If we use a value for of 528.29 K for $T_F$ in the expression for S, this results in a Seebeck value of −28.9 μV/K at 35K, which yields excellent matching with the measured value of −30.5 μV/K at the same temperature (relative to the value of copper $S_{Cu} \sim 1.8$ μV/K). That signifies that the effective mass of electron equals to its rest mass around 35 K. This strongly underlines the applicability of Fermi liquid theory in this temperature regime from the zero field resistivity analysis. The Sommerfeld coefficient as given for free electron model[34] is $\gamma = \pi^2 N k_B / 2T_F$ where N is the total charge carriers calculated for one mole of molecule. $Bi_4O_4S_3$ is the normalized formula for 50% $SO_4$ deficient $Bi_6O_8S_5$ so taking the molecular weight of $Bi_6O_8S_5$-0.5$SO_4$ or $Bi_6O_6S_{4.5}$ and using the calculated density 7.62 gm/cm$^3$, the Sommerfeld constant comes out to be ~ 1.113 mJ/K$^2$ mol. It needs to be pointed out that our estimation of Sommerfeld coefficient is smaller than the value reported by Hiroshi et al from specific heat measurements which is ~2.8 mJ/K$^2$ mol.[5]

To test the potentiality of $BiS_2$ based superconductors towards applications, in the following we discuss the basic pinning and vortex activation characteristics of $Bi_4O_4S_3$. The broadening of the resistive transition under applied magnetic field is understood in terms of a dissipation of energy caused by the motion of vortices in the mixed state. This interpretation is based on the fact that for the low-resistance region, the lossy behavior is caused by the creep of vortices which is modeled as Arhenius type thermally activated behavior. This is simply described by $\rho(T, B) = \rho_0 \exp(-U_0/k_BT)$. Here, $U_0$ is the flux-flow activation energy, that can be obtained from the slope of the linear parts of an Arrhenius plot and $\rho_0$ is a field-independent pre-exponential factor. Investigations of high-Tc superconductors and artificial multilayers have showed that the activation energy exhibits different power-law dependences on a magnetic field, i.e. $U_0(H) \propto H^{-n}$.[36-38] The values of $U_0$ (Figure 8), are deduced from the limited temperature intervals below $T_c$, in which the data of the Arrhenius plot of $\rho(T)$ yield straight lines. The Arrhenius plots are shown in Figure 8 and activation energy in inset. The straight-line behavior over 5 decades of the resistivity data validates the thermally activated flux flow (TAFF)-process as described by the Arrhenius law. The best fit of the experimental data $\rho(T)|_{B=const}$ yields values of the activation energy, ranging from $U_0/k_B$ ~ 121 K in low magnetic field (0.01T) down to ~ 9.75 K in the high field region (1T) (Inset Fig. 8). The value of the exponent n in the power law equation $U_0(B) \sim B^{-n}$ is ~0.3 for H < 0.1. On the other hand, $Bi_4O_4S_3$ shows a much stronger field dependence of the activation energy in the high magnetic field region with the value of n ~1.99 for H > 0.1T.

## Conclusions

In conclusion, we have successfully synthesized a series of $BiS_2$ layer based $Bi_6O_4S_4(SO_4)_{1-x}$ superconductors with varying $SO_4$ concentration. The optimally doped superconducting sample $Bi_4O_4S_3$ has transition temperature 5.3 K which is the highest $T_c$ reported for this compound. The crystal structure establishes the increase in the c-axis for superconducting sample to that of the parent $Bi_6O_8S_5$ that relates to the bending of Bi2-S2-Bi2 bond in the $BiS_2$ layer. The upper critical field was estimated ~2.75T for $\rho_{onset}$ and zero temperature coherence length $\xi_{GL}(0)$ ~ 110 Å for $H_{c2}(\rho_{onset})$. The pinning potential $U_0$ scales as $B^{-0.3}$ for B < 0.1T. Hall measurements confirm dominance of electronic conduction and in conjuction with thermoelectric power data, the Sommerfeld's constant is estimated to be ~ 1.113 mJ/$K^2$ mol. The energy gap calculated through rf susceptibility establishes this system to be strongly coupled superconductor and the data fits well for an isotropic S wave gap.

## Acknowledgment

PS is grateful to UGC, India for the award of a Dr D S Kothari postdoctoral fellowship. Shruti acknowledge the UGC for providing her the JRF fellowship. Technical support from AIRF is gratefully acknowledged. This work is performed under the sponsorship of DST-FIST.

**Table Caption**

**Table 1.** Structural and lattice parameters of $Bi_6O_8S_5$ and $Bi_4O_4S_3$ obtained by Rietveld refinement. Bi1, Bi2, Bi3, S1, and S2 occupy the 4e (0, 0, z) site. The S3 atom is at the 2b (0, 0, 1/2) site. O1 is situated at the 8g (0, 1/2, z) site, and O2 is positioned at the 16n (0, y, z) site. Only the variable z coordinates are given for Bi and S atoms in the table while the O1 and O2 are fixed at (0, 0.5, 0.0884) and (0, 0.3053, 0.4793) respectively [1].

**Figure captions**

**FIGURE 1.** (a) Schematic representation of the crystal structure for $Bi_4O_4S_3$. Grey, yellow and red circles indicate Bi, S and O atoms, respectively. The structure is composed of $Bi_2O_2$, $BiS_2$ and $SO_4$ layers. (b) $BiS_2$ layer with out of plane angles and bond lengths. (c) Image of the $BiS_2$ square lattice (*ab* plane) with in plane parametres.

**FIGURE 2.** XRD pattern with the result of the Rietveld refinement. Bragg reflections for $Bi_4O_4S_3$, $Bi_2S_3$ and Bi are shown by green vertical lines.

**FIGURE 3.** Scanning Electron Microscopy (SEM) images show the morphology of (a) $Bi_6O_8S_5$ and (b) $Bi_4O_4S_3$ samples. The average length and breadth of $Bi_4O_4S_3$ crystals is ~ 6 μm and ~1 μm respectively.

**FIGURE 4** (a). Temperature dependence of resistivity for $Bi_4O_4S_3$. Solid curve shows the fitting of $A+BT^n$ with n=0.78. Inset (i) shows semiconducting resistivity profile for $Bi_6O_8S_5$ (ii) Enlarged low-temperature data, with the indicated superconducting transition temperature (iii) Change in transition temperature with doping concentration x for of $Bi_3O_{4-4x}S_{2.5-x}$ (where x=0, 0.2, 0.25, 0.4, 0.5, 0.6, 0.66, 0.75, 0.8).
(b) $\rho = \rho_0 + AT^2$fitiing with value of $\rho_0$ = 0.847 mΩ-cm and the coefficient $A = 3.12\times10^{-5}$ mΩ-cm/$K^2$. Inset (i) enlarged view of the fitting from 25K to 50K. (ii) Change in frequency with respect to temperature.

**FIGURE 5.** (a) Resistivity vs temperature (ρ−T) behavior of $Bi_4O_4S_3$ in applied fields of 0, 0.01, 0.02, 0.03, 0.05, 0.1, 0.25, 0.5, 0.8, 1, 2, 3 and 4T in the superconducting region. Inset

shows the persistent kink in the resistivity at higher magnetic fields of 2T, 3T, 4T. (b) Represent the GL approach extrapolated to zero temperature, Inset shows the upper critical field $H_{c2}$ and the $H_{irr}$ found from onset and offset temperature.

**FIGURE 6.** Hall resistivity as a function of magnetic field at 10 K and 300K. Insets shows magnetoresistance at 6K as a function of square of the magnetic field.

**FIGURE 7.** Temperature dependence of Seebeck coefficient for $Bi_4O_4S_3$ in zero magnetic field. The Seebeck coefficient is negative in the whole temperature range with maximum negative value of ~30.5 μV/K.

**FIGURE 8.** (a) Arrhenius plot of the resistivity for $Bi_4O_4S_3$ for (left to right) with H = 0, 0.01, 0.02, 0.03, 0.05, 0.1, 0.25, 0.5, 0.8, 1, 2, 3 and 4T. (Inset) Dependence of the activation energy $U_0/k_B$ on magnetic field for the samples.

**Table 1.**

| Parameters | $Bi_6O_8S_5$ | $Bi_4O_4S_3$ |
|---|---|---|
| **a (Å)** | 3.9655 | 3.9675 |
| **c (Å)** | 41.2340 | 41.3228 |
| **V ( $Å^3$)** | 648.41 | 650.46 |
| **z Bi1**<br>**Bi2**<br>**Bi3** | 0.0612(8)<br>0.2078(11)<br>0.3767(9) | 0.0601(2)<br>0.2084(0)<br>0.3815(8) |
| **z S1**<br>**S2** | 0.127(5)<br>0.287(6) | 0.1313(6)<br>0.2811(4) |
| **Bi2-S1 (Å)** | --- | 3.18(8) |
| **S1-Bi2-S2** | 180° | 180° |
| **(In plane) Bi2-S2 (Å)**<br>**(out of plane) Bi2-S2 (Å)** | 2.812(19)<br>2.812(19) | 2.839(13)<br>3.00(8) |
| **(In plane) Bi2-S2-Bi2**<br>**(out of plane) Bi2-S2-Bi2** | 89.7°<br>171.3° | 88.6°<br>162.0° |

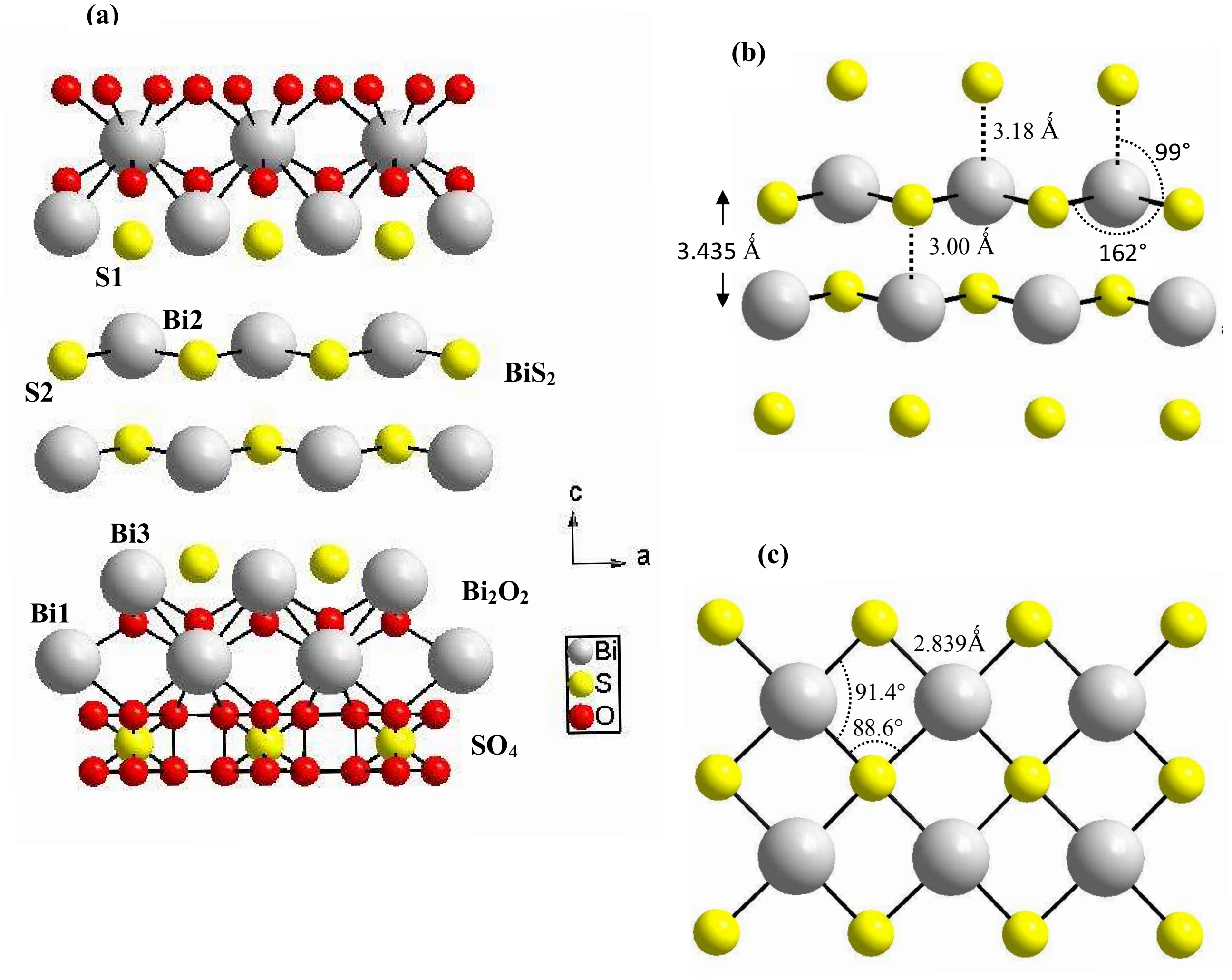

**FIGURE 1.**

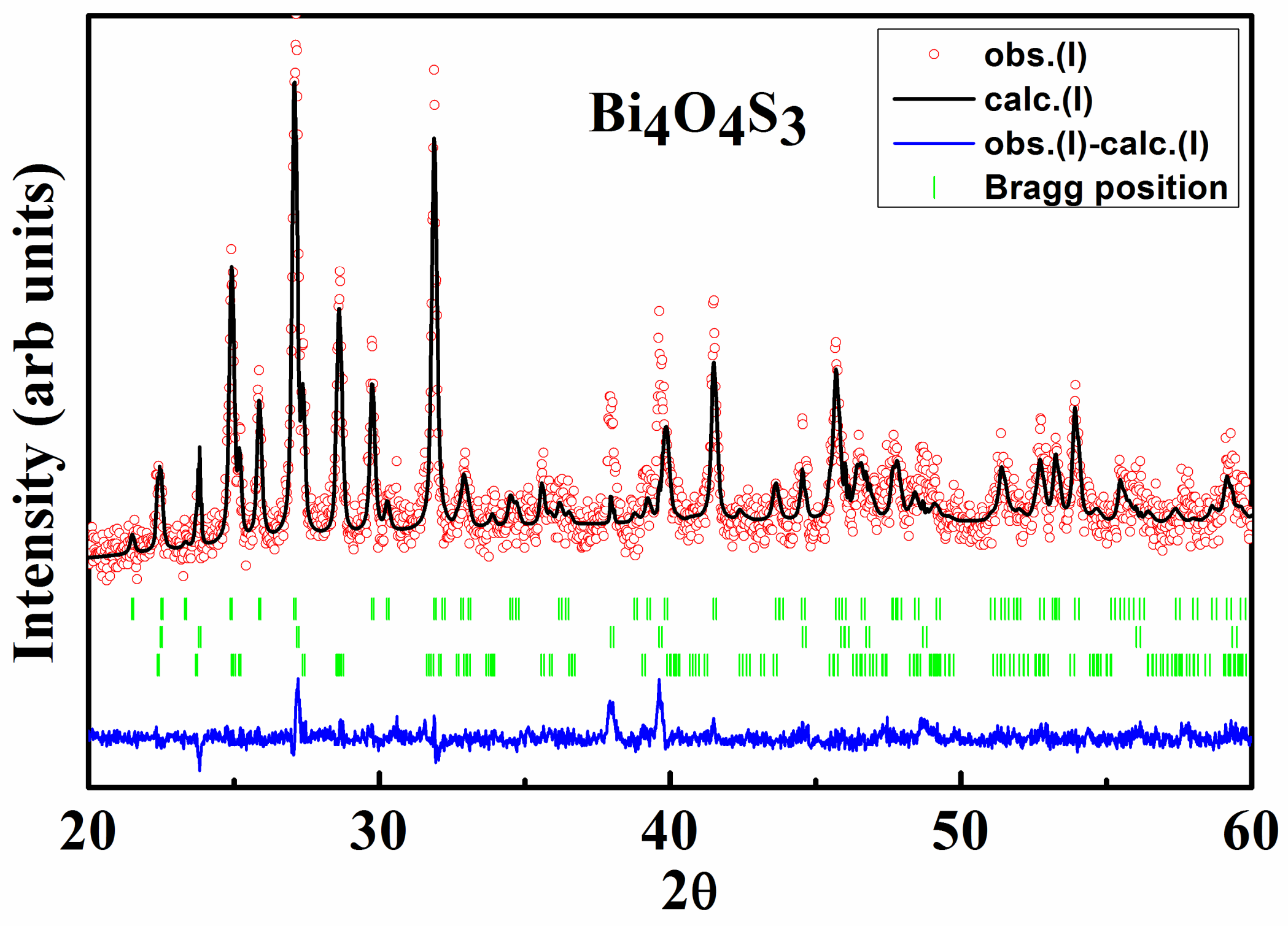

**FIGURE 2.**

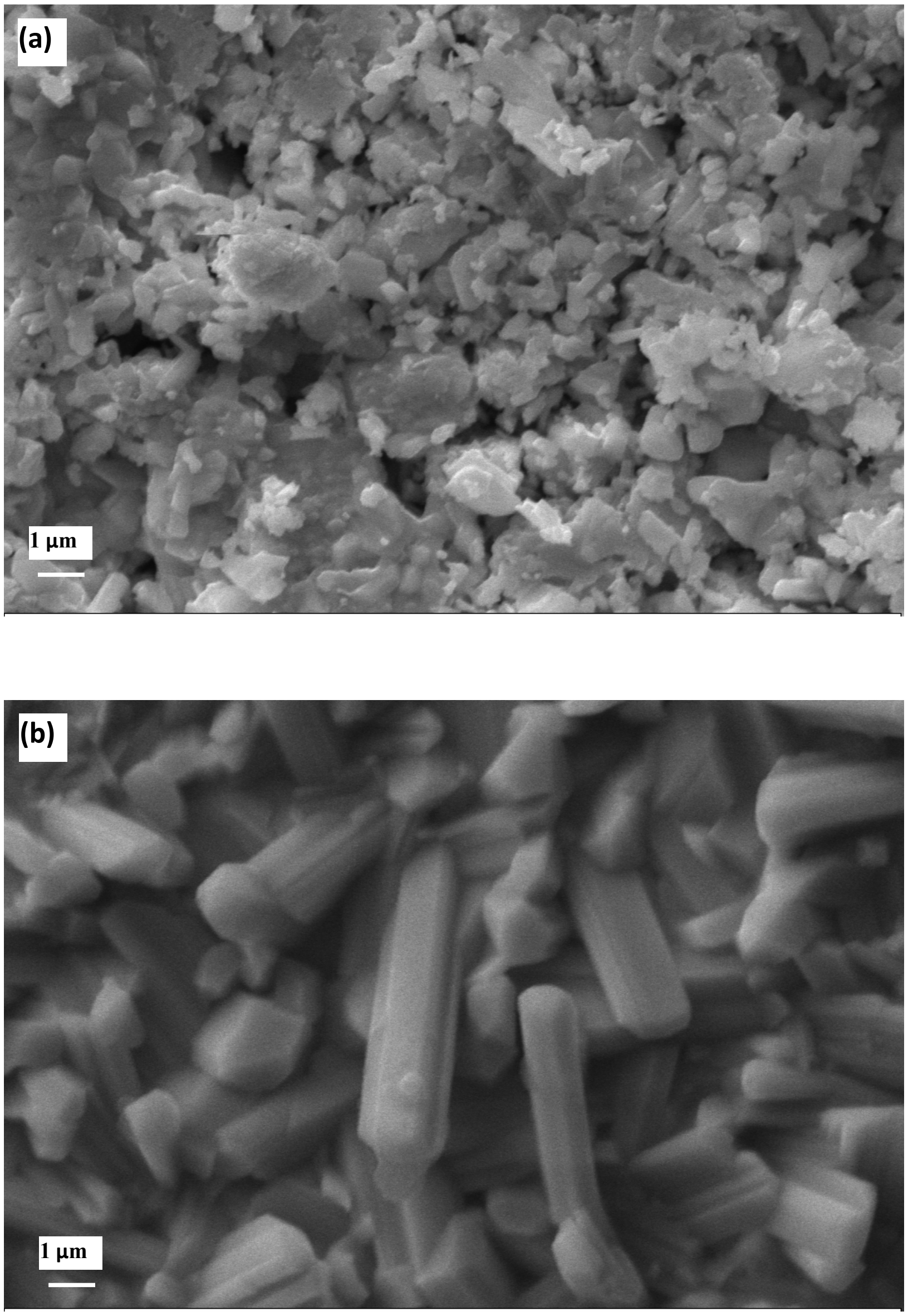

**FIGURE 3.**

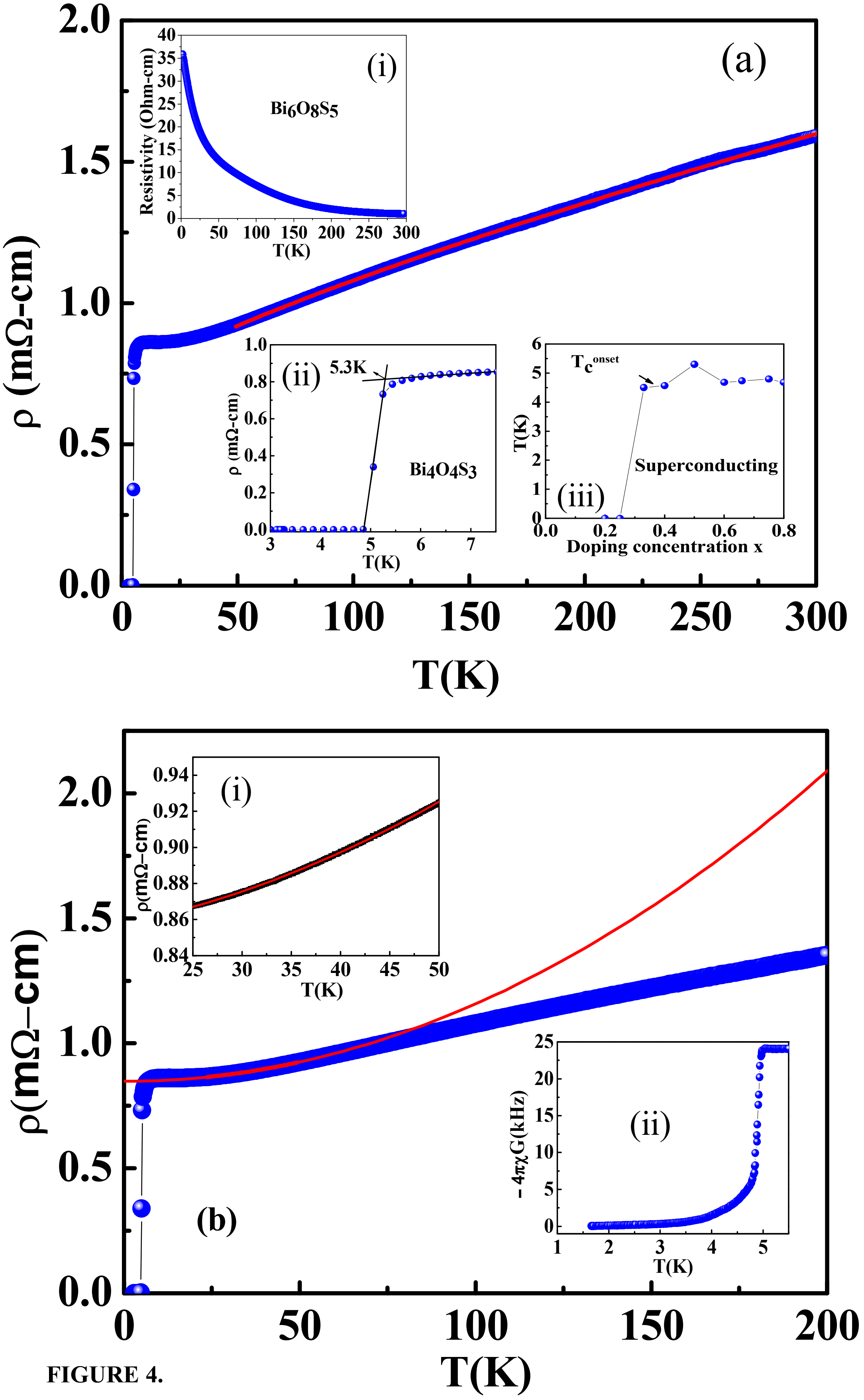

FIGURE 4.

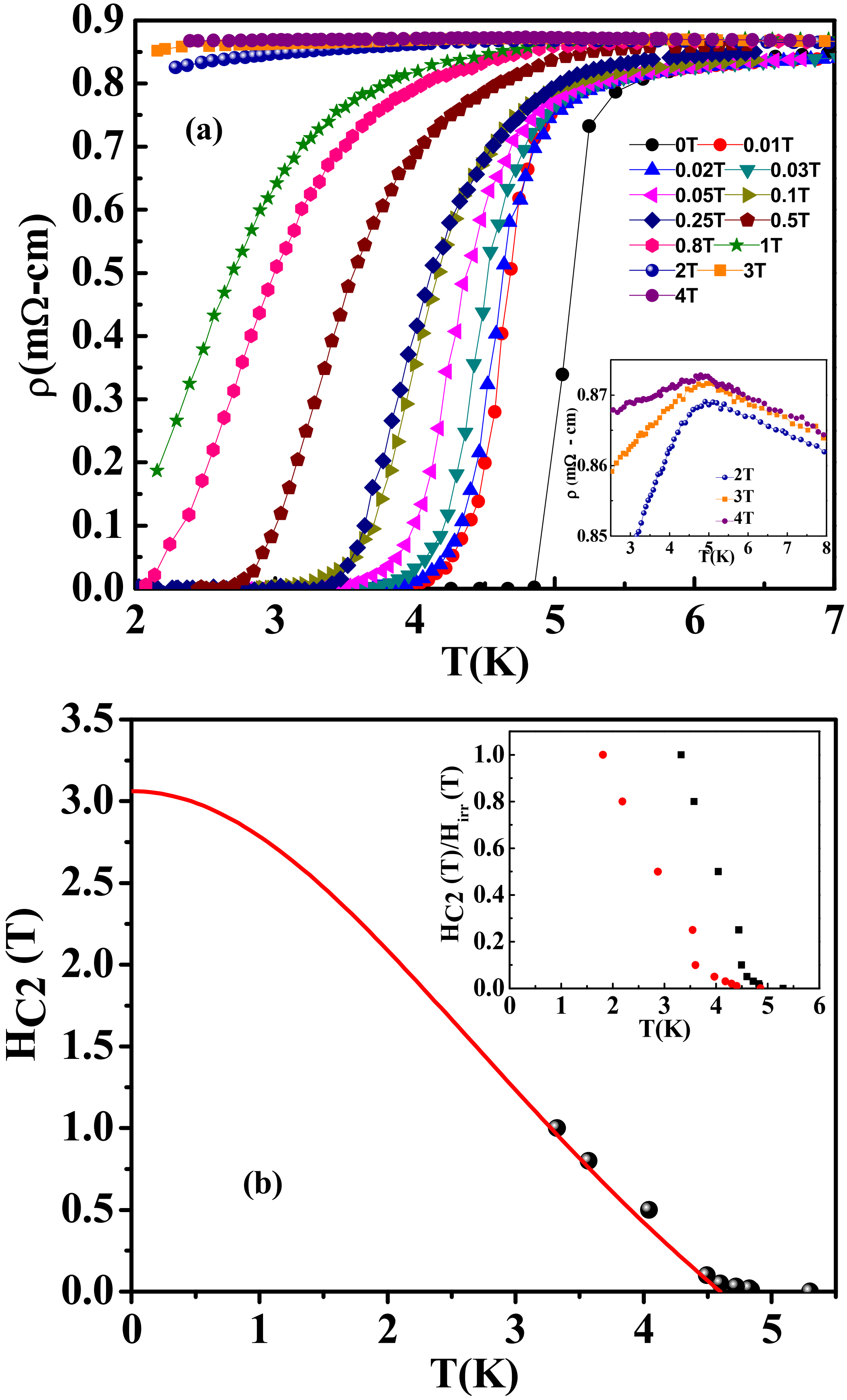

FIGURE 5.

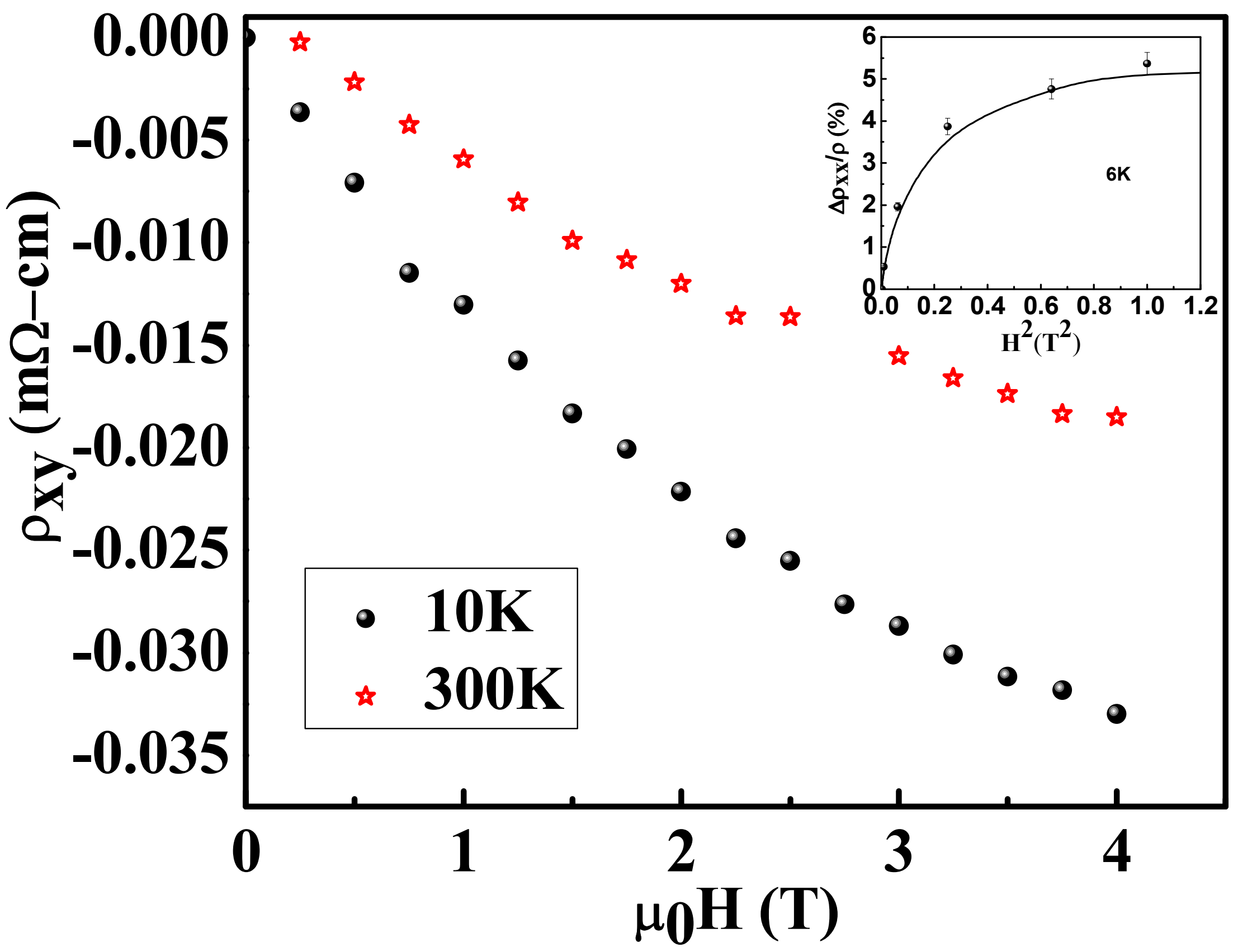

**FIGURE 6.**

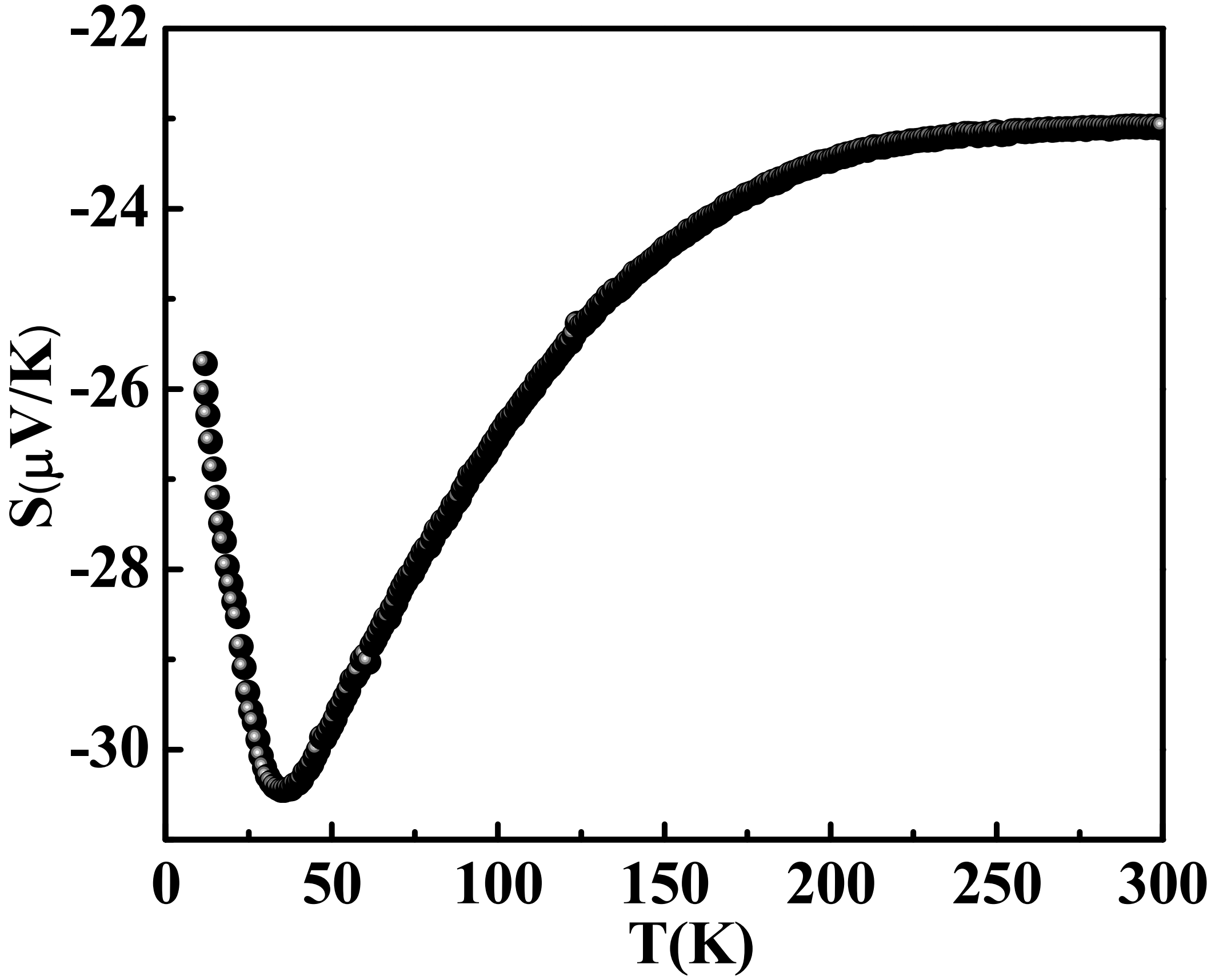

**FIGURE 7.**

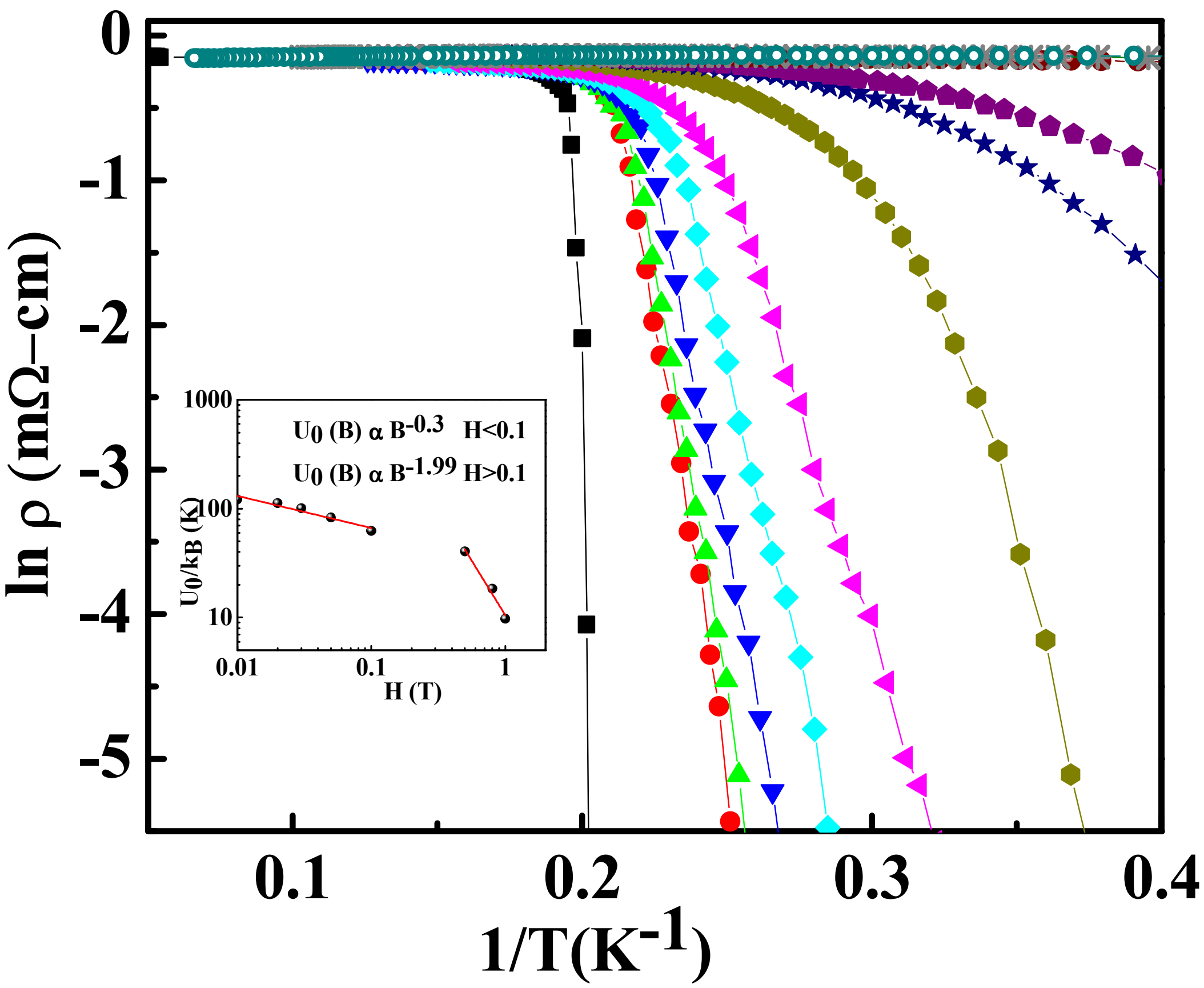

FIGURE 8.